\documentclass[%
article,letterpaper,
superscriptaddress,
 amsmath,amssymb,
 aps,
]{revtex4}
\usepackage{tocvsec2}
\usepackage[utf8]{inputenc}
\usepackage{hyperref}
\hypersetup{
    colorlinks=true,
    linkcolor=blue,
    filecolor=magenta,      
    urlcolor=magenta,
    citecolor=blue
}

\usepackage[top=1in, bottom=1.25in, left=1.25in, right=1.25in]{geometry}
\usepackage{titlesec}
\usepackage{enumerate}
\usepackage{enumitem}
\usepackage{bm}
\usepackage{amsfonts}
\usepackage{amsmath}
\usepackage{amssymb}
\usepackage{etoolbox}
\newcommand{\nubar}[0]{\overline{\nu}}
\patchcmd{\thebibliography}{\chapter*}{\section*}{}{}

\titlespacing*{\chapter}{0pt}{-50pt}{20pt}
\titleformat{\chapter}[display]{\normalfont\large\bfseries}{\chaptertitlename\ \thechapter}{20pt}{\Large}
\titleformat*{\section}{\fontsize{16}{20}\selectfont}

\begin{document}
\title{Snowmass 2021 LoI:

Neutrino-induced Shallow- and Deep-Inelastic Scattering 

}
%
\author{L. Alvarez-Ruso}
\affiliation{Instituto de F\'isica Corpuscular (IFIC), Centro Mixto Universidad de Valencia-CSIC,
Institutos de Investigaci\'on de Paterna, Apartado 22085, 46071 Valencia, Spain}

\author{A. M. Ankowski}
\affiliation{SLAC National Accelerator Laboratory, Stanford University, Menlo Park, CA 94025, USA}

\author{M.~Sajjad Athar} 
\affiliation{AMU Campus, Aligarh, Uttar Pradesh 202001, India}

\author{C.~Bronner} \affiliation{Kamioka Observatory, Institute for Cosmic Ray Research, University of Tokyo, Kamioka, Gifu, Japan}

\author{L.~Cremonesi}
\affiliation{Queen Mary University of London, London E1 4NS, UK}

\author{K.~Duffy}
\affiliation{Fermi National Accelerator Laboratory, Batavia, IL 60510, USA}

\author{S.~Dytman} 
\affiliation{University of Pittsburgh, Pittsburgh, PA, 15260, USA}

\author{A. Friedland}
\affiliation{SLAC National Accelerator Laboratory, Stanford University, Menlo Park, CA 94025, USA}

\author{A. P. Furmanski}
\affiliation{University of Minnesota, Twin Cities, Minneapolis, MN, 55455, USA}

\author{K.~Gallmeister}
\affiliation{Institut für Theoretische Physik, Goethe-Universität Frankfurt, Frankfurt am Main, Germany}

\author{S.~Gardiner}
\affiliation{Fermi National Accelerator Laboratory, Batavia, IL 60510, USA}

\author{W.~T.~Giele}
\affiliation{Fermi National Accelerator Laboratory, Batavia, IL 60510, USA}

\author{N.~Jachowicz} 
\affiliation{Department of Physics and Astronomy, Ghent University, B-9000 Gent, Belgium}

\author{H.~Haider} 
\affiliation{AMU Campus, Aligarh, Uttar Pradesh 202001, India}

\author{M.~Kabirnezhad} 
\affiliation{University of Oxford, Oxford OX1 3RH, United Kingdom}

\author{T.~Katori} 
\affiliation{King's College London, London WC2R 2LS, UK}

\author{A.~S.~Kronfeld}
\affiliation{Fermi National Accelerator Laboratory, Batavia, IL 60510, USA}

\author{S.~W.~Li} 
\affiliation{SLAC National Accelerator Laboratory, Stanford University, Menlo Park, CA 94025, USA}

\author{J.G.~Morf\'{i}n} 
\affiliation{Fermi National Accelerator Laboratory, Batavia, IL 60510, USA}

\author{U.~Mosel}
\affiliation{Institut für Theoretische Physik, Universität Giessen, Giessen, Germany}

\author{M.~Muether} 
\affiliation{Wichita State University, Wichita, KS 67260, USA}

\author{A.~Norrick}
\affiliation{Fermi National Accelerator Laboratory, Batavia, IL 60510, USA}

\author{J.~Paley}
\affiliation{Fermi National Accelerator Laboratory, Batavia, IL 60510, USA}

\author{V.~Pandey}
\affiliation{Department of Physics, University of Florida, Gainesville, FL 32611, USA}

\author{R.~Petti}
\affiliation{Department of Physics and Astronomy, University of South Carolina, Columbia SC 29208, USA}

\author{L. Pickering}
\affiliation{Department of Physics and Astronomy, Michigan State University, East Lansing MI 48824, USA}

\author{B. J. Ramson}
\affiliation{Fermi National Accelerator Laboratory, Batavia, IL 60510, USA}

\author{M. H. Reno}
\affiliation{Department of Physics and Astronomy, University of Iowa, Iowa City, IA, 52242, USA}

\author{T. Sato}
\affiliation{Department of Physics, Osaka University, Osaka 560-0043, Japan}

\author{J.T. Sobczyk}
\affiliation{Institute of Theoretical Physics, Wroc\l aw University, 50-204 Wroc\l aw, Poland}

\author{J.~Wolcott}
\affiliation{Department of Physics and Astronomy, Tufts University, Medford, MA, 02155, USA}

\author{C.~Wret}
\affiliation{Department of Physics and Astronomy, University of Rochester, Rochester, New York, 14627, USA}

\author{T.~Yang}
\affiliation{Fermi National Accelerator Laboratory, Batavia, IL 60510, USA}

\begin{abstract}
\end{abstract}

\maketitle

\noindent {\large \bf NF Topical Groups:}  (check all that apply $\square$/$\blacksquare$)

\noindent $\blacksquare$ (NF1) Neutrino oscillations\\
\noindent $\square$ (NF2) Sterile neutrinos \\ 
\noindent $\square$ (NF3) Beyond the Standard Model  \\
\noindent $\square$ (NF4) Neutrinos from natural sources\\
\noindent $\square$ (NF5) Neutrino properties \\
\noindent $\blacksquare$ (NF6) Neutrino cross sections \\
\noindent $\square$ (NF7) Applications \\
\noindent $\blacksquare$ (TF11) Theory of neutrino physics \\
\noindent $\square$ (NF9) Artificial neutrino sources \\
\noindent $\blacksquare$ (NF10) Neutrino detectors  \\
\noindent $\blacksquare$ (Other) { CompF2 (Theoretical Calculations and Simulation)} \\

\noindent {\bf Contact Information:}
Teppei Katori, \href{mailto:teppei.katori@kcl.ac.uk}
{teppei.katori@kcl.ac.uk}

\newpage 
\noindent{\bf Introduction} --- 
In $\nu/\nubar$ interactions with nucleons and nuclei Shallow Inelastic Scattering (SIS) refers to processes, dominated by non-resonant contributions, in the kinematic region where $Q^2$ is small and the invariant mass of the hadronic system, $W$, is above pion production threshold. As $W$ increases above the baryon-resonance dominated region, non-resonant meson production begins to play a significant role. In addition, as $Q^2$ grows, one approaches the onset of the DIS region. The extremely rich science of this complex region, poorly understood both theoretically and experimentally~\cite{SajjadAthar:2020nvy, Alvarez-Ruso:2017oui,Andreopoulos:2019gvw}, encompasses the \emph{transition} from interactions described in terms of hadronic degrees of freedom to  interactions with quarks and gluons described by perturbative QCD. Neutrino-nucleus experiments cannot distinguish mesons from these contributing processes, thus the experimental definition of SIS is the inclusive sum of these processes in the higher W region bordering DIS. Since a large fraction of events in NOvA~\cite{Acero:2019ksn} and DUNE~\cite{Abi:2020wmh}, and in atmospheric neutrino measurements at IceCube-Upgrade~\cite{Ma:2020wmj}, KM3NeT~\cite{Adrian-Martinez:2016fdl}, Super- and Hyper-Kamiokande~\cite{Fukuda:2002uc,Abe:2018uyc}, are from this SIS region, there is a definite need to improve our knowledge of this physics. 

\vspace{0.2cm}
\noindent{\bf Inelastic processes} --- Neutrino-nucleon inelastic scattering predominantly leads to single pion ($\pi N$) but also to $\gamma N$, $\pi \pi N$, $\eta N$, $\rho N$, $K N$, $\bar{K} N$ $K Y$, ... final states. Close to threshold, elementary amplitudes are constrained by the approximate chiral symmetry of QCD \cite{Hernandez:2007qq, RafiAlam:2010kf, Alam:2012zz,Yao:2018pzc}. Away from threshold, most of these reactions are dominated by baryon resonances, albeit with sizable contribution from non-resonant amplitudes and their interference with the resonant counterpart. This is the case for the $\Delta(1232)$-dominated single pion production \cite{Hernandez:2007qq}, for which extensions to higher invariant masses within the Regge approach have also been developed~\cite{Gonzalez-Jimenez:2016qqq,Gonzalez-Jimenez:2017fea,Nikolakopoulos:2018gtf}. The Rein-Sehgal model~\cite{Rein:1980wg} is an outdated model for inelastic processes which, nevertheless, is still widely used in event generators \cite{Kabirnezhad:2017jmf}.  The dynamical coupled channel (DCC) approach~\cite{Nakamura:2015rta,Kamano:2013iva,Kamano:2016bgm} is consistently constrained using $eN$ and $\pi N$ vast amount of  data to predict not only weak single but also double pion production and other meson-baryon final states up to $W$ of about 2.2 GeV. Indeed, thanks to flavor symmetries and the partial conservation of the axial current (PCAC), electron- and meson-nucleon scattering  provide very valuable input for the description of inelastic processes but the axial current remains largely unconstrained.
The Giessen BUU model~\cite{Buss:2011mx} relies on the MAID analysis~\cite{Drechsel:2007if} of electron-nucleon pion production and PCAC to constrain elementary amplitudes, and on transport theory to model the evolution of the final state to describe exclusive channels in neutrino-nucleus inelastic scattering.  

\vspace{0.2cm}
\noindent{\bf Quark-Hadron Duality} ---
The transition from resonant/non-resonant production to DIS is marked by increasing $W$, which in turn with growing $Q^2$, naturally evolves into scattering off the quark in the nucleon that can be described by perturbative QCD.  On the way to this QCD-described scattering region there is a significant contribution from the  non-perturbative QCD regime. This is a very complex kinematic transition region, encompassing interactions that can be described in terms of hadrons as well as quarks, that should be well-described by the application of quark-hadron duality~\cite{Bloom:1970xb} where baryonic resonant and non-resonant processes behave on average like DIS in similar $Q^2$ and $W$ regions.  Although duality has been demonstrated with electromagnetic  induced processes, it has neither been well studied theoretically nor are there experimental results in the weak sector. More experimental and theoretical studies are required to understand this intriguing region~\cite{SajjadAthar:2020nvy}. 

\vspace{0.2cm}
\noindent{\bf DIS in the Nuclear Environment} ---
The investigation of DIS within the nucleus in the electromagnetic sector revealed that nuclear effects modify structure functions and consequently nuclear parton distribution functions (nPDFs) are different than nucleon PDFs. Theory has indicated how nuclear effects could modify nucleon structure functions to yield nuclear structure functions~\cite{Kulagin:2007ju,Kulagin:2004ie,Kulagin:2014vsa} while phenomenological global fits have directly yielded “effective” nuclear PDFs~\cite{Schienbein:2009kk,Kovarik:2010uv}. “Effective” since, within the nuclear environment, scattering could be occurring with more than a single hadron that need not even be a nucleon.  More recent investigations of neutrino-nucleus scattering have suggested that the resultant nPDFs could be different from those derived from electromagnetic scattering~\cite{Haider:2016zrk, Zaidi:2019asc,Tice:2014pgu,Mousseau:2016snl}.  These differences need to be further explored.

\vspace{0.2cm}
\noindent{\bf Hadronization} ---
Hadronization is not described by a fundamental theory, but based on phenomenological models. At low $W$ ($W\lesssim 3$~GeV) neutrino interaction generators have to use custom models based on empirical KNO scaling~\cite{KOBA1972317}, with parameters tuned to old bubble chamber data~\cite{Yang:2009zx, doi:10.7566/JPSCP.12.010041}. These data often lack systematic uncertainties, and lead to  inconsistent results~\cite{PhysRevC.88.065501,Katori:2014fxa}: the description of hadronization would benefit from new data, taken in modern experiments~\cite{Chukanov:2016lra,Adams:2018fud,Hiramoto:2020gup}. As $W$ increases (from $W$ of 2 to 3 GeV), the hadronic system becomes too complex to use custom models, and the generators rely on the models built for colliders~\cite{Sjostrand:2006za,Corcella:2000bw,Gleisberg:2008ta}. Although these models could be successfully built to simulate hadronization in high-energy collider experiments, these $W$ values at the SIS region is lower than the validity range of the models; additional developments would be therefore needed. 

\vspace{-0.4cm}
\subsection*{Path forward}
\vspace{-0.2cm}
\noindent Current and future oscillation experiments need a better understanding and realistic modeling of neutrino-nucleus SIS scattering. To meet this challenge, we need coordinated work by both nuclear physics and particle physics communities; in theory, experiment, and simulation. Such a commitment is beneficial to both communities to achieve broader scientific goals in multidisciplinary topics. 

\vspace{0.2cm}
\noindent{\bf Theoretical challenges} ---
Realistic theoretical modeling of SIS scattering should provide accurate predictions of neutrino-nucleus interactions, as well as meaningful theoretical uncertainties. This can only be achieved if existing and future neutrino scattering data on nucleons and nuclei but also electron and pion-nucleon scattering data are systematically incorporated both as input and for model validation. In the resonance-dominated region, one also aims at a description 
in terms of well-defined final states 
where resonant and non-resonant terms and their interferences are consistently treated. In the transition from SIS to DIS, differences between Monte Carlo generators often yield inconsistent predictions, as shown graphically in, e.g., \cite{SajjadAthar:2020nvy} and by Bronner in \cite{Andreopoulos:2019gvw}. The pioneering PDF-based approach of Bodek-Yang \cite{Yang:1998zb,Bodek:2002ps,Bodek:2004pc,Bodek:2010km} and more phenomenological, theory-guided structure function approaches that do not rely on a parton decomposition (see, e.g., \cite{Capella:1994cr,Reno:2006hj}), merit study in view of the availability of more recent PDFs, studies of target mass and higher twist corrections, and next-to-next-to-leader order \cite{Vermaseren:2005qc,Moch:2004xu,Moch:2008fj} perturbative treatments of DIS \cite{Zaidi:2019asc}. 

\vspace{0.2cm}
\noindent{\bf Experimental challenges} --- We identify especially three categories to improve our knowledge. 

{\it Neutrino-hydrogen/deuteron scattering experiments} --- 
Even if electron- and meson-nucleon scattering data provide a priceless input to model neutrino interactions on nucleons \cite{Nakamura:2015rta,Kabirnezhad:2020wtp}, the properties of the axial current at finite $Q^2$ remain largely unknown and experimentally unconstrained.  
For example, most axial form factors are not directly measured. Although lattice QCD may be able to partially fill this gap
~\cite{Cirigliano:2019jig},  
we need modern neutrino-hydrogen and/or neutrino-deuteron scattering experiment to directly measure unknown form factors (see $\nu$-H/D LoI~\cite{Snowmass2021:nu-H}). 

{\it Electron-nucleus scattering experiments} --- Modern neutrino-nucleus models use electron-nucleus vector interaction model, but axial interactions poorly constrained by neutrino-nucleon data must be added.  Precision measurements of inclusive electron-nucleus scattering at a wide variety of kinematics are important for validating nuclear models~\cite{Ankowski:2020qbe}. 
Recent electron scattering measurements~\cite{Dai:2018xhi,Dai:2018gch,adi_ashkenazi_2020_3959538} on various targets (including Ar) indicate moderate discrepancies in the generator models beyond the quasielastic peak~\cite{Papadopolou:2020zkd}. Coverage must be extended into the SIS kinematic region and information on the final-state mesons and nucleons should be added~\cite{e4nu17,e4nu18,Ankowski:2019mfd} so that FSI models can be tested. 

{\it Neutrino-nucleus scattering experiments} --- 
Neutrino oscillation experiments 
such as MiniBooNE, T2K, NOvA, and MINERvA have published cross-section data mainly for $\text{CH}_n$ and $\text{H}_2\text{O}$ targets in QE region. Limited data on heavier targets (Ar, Fe, Pb) and higher energy processes are also available~\cite{Tice:2014pgu,Mousseau:2016snl,Betancourt:2017uso,Abe:2015biq,Hiramoto:2020gup,Adamson:2016hyz,Wu:2007ab,Adamson:2009ju,Lyubushkin:2008pe}. These data offer an opportunity to test nuclear dependent DIS models in neutrinos. 
The SBN program (MicroBooNE, SBND, ICARUS)~\cite{Antonello:2015lea} and ArgonCube~\cite{Abi:2020wmh} can provide Ar cross-section data relevant for the SIS region.  
More extensive experimental studies focusing on meson final states in a broad kinematic range can test our understanding of the neutrino SIS physics as well as FSIs~\cite{PhysRevD.100.072005}. 

\vspace{0.2cm}
\noindent{\bf Generator challenges} ---
With the lack of a coherent picture of the SIS region, 
the models presently used in generators are either smoothed descriptions of inclusive data or often inconsistent mixtures of models \cite{Mosel:2019vhx}. 
Recently, a fairly complete group of generator experts started a new initiative to
improve structural issues~\cite{Barrow:2020gzb}. The present task to develop a consistent and accurate SIS model is a very interesting and challenging physics problem that requires proficiency in both nuclear physics and particle physics.  One of the sources of the present inconsistency is the different framing in different sub-fields.  A more complete picture is needed to achieve a coherent model. (see $\nu$-generator LoI~\cite{Snowmass2021:nu-generator})

\newpage 

\bibliographystyle{unsrt}
\bibliography{biblio}

\end{document}